# ESM-NBR: fast and accurate nucleic acid-binding residue prediction via protein language model feature representation and multi-task learning

Wenwu Zeng[1]
[1] College of Computer Science and Electronic Engineering
Hunan University
Changsha, China
wwz_cs@126.com

Dafeng Lv [1]
[1] College of Computer Science and Electronic Engineering
Hunan University
Changsha, China
lvdafeng2001@163.com

Wenjuan Liu [1,*]
[1] College of Computer Science and Electronic Engineering
Hunan University
Changsha, China
liuwenjuan89@hnu.edu.cn

Shaoliang Peng [1,*]
[1] College of Computer Science and Electronic Engineering
Hunan University
Changsha, China
slpeng@hnu.edu.cn

*Abstract*—Protein-nucleic acid interactions play a very important role in a variety of biological activities. Accurate identification of nucleic acid-binding residues is a critical step in understanding the interaction mechanisms. Although many computationally based methods have been developed to predict nucleic acid-binding residues, challenges remain. In this study, a fast and accurate sequence-based method, called ESM-NBR, is proposed. In ESM-NBR, we first use the large protein language model ESM2 to extract discriminative biological properties feature representation from protein primary sequences; then, a multi-task deep learning model composed of stacked bidirectional long short-term memory (BiLSTM) and multi-layer perceptron (MLP) networks is employed to explore common and private information of DNA- and RNA-binding residues with ESM2 feature as input. Experimental results on benchmark data sets demonstrate that the prediction performance of ESM2 feature representation comprehensively outperforms evolutionary information-based hidden Markov model (HMM) features. Meanwhile, the ESM-NBR obtains the MCC values for DNA-binding residues prediction of 0.427 and 0.391 on two independent test sets, which are 18.61 and 10.45% higher than those of the second-best methods, respectively. Moreover, by completely discarding the time-cost multiple sequence alignment process, the prediction speed of ESM-NBR far exceeds that of existing methods (5.52s for a protein sequence of length 500, which is about 16 times faster than the second-fastest method). A user-friendly standalone package and the data of ESM-NBR are freely available for academic use at: https://github.com/pengsl-lab/ESM-NBR.

*Keywords—nucleic acid-binding residue prediction; multi-task learning; protein language model feature representation*

## I. Introduction

Interactions between proteins and nucleic acids (DNA/RNA) play an indispensable role in many biological processes, e.g., DNA replication, transcription, recombination, protein synthesis, regulation of gene expression, and transcriptional modifications [1-4]. Accurate identification of nucleic acids-binding residue is one of most key steps in understanding protein-nucleic acids interaction. The mainstream methods for determining nucleic acids-binding residue are primarily based on wet-lab experimental like nuclear magnetic resonance (NMR), chromatin immunoprecipitation on the microarray (ChIP-chip), genetic analysis [5], and X-ray crystallography [6]. These methods study protein-nucleic acids interactions by determining complex structure and have made great contribution. However, due to the inherent limitations of wet-lab experiment (time-consuming, laborious, and costly), there are only 6,268 protein-DNA complexes and 2,838 protein-RNA complexes have been resolved in the Nucleic Acid Database [7] until June 30, 2023, respectively. This is a huge gap with the number of known nucleic acid-binding proteins [8]. Although the recently proposed RoseTTAFoldNA [9] made some progress in predicting nucleic acid complexes, they also acknowledge that there is still great room for further improvement the accuracy. In this context, with the rapid development of sequencing technology in the post-genetic era [10], there is an urgent demand for a high throughput and accurate computation-based method to identify nucleic acids-binding residue.

Over the past few decades, researchers have developed a series of computational models based known experimental data for predicting nucleic acids-binding residue. Depending on the input information they used, these methods can be roughly divided into two categories: structure-based, e.g., GraphSite [11] (predicted structure from AlphaFold2), PST-PRNA [12], GraphBind [13], PRIME-3D2D [14], and COACH-D [15], and sequence-based, e.g., ProNA2020 [16], DNAPred [17], iDRNA-ITF [18], Pprint2 [19], NCBRPred [20], and DRNApred [8]. The former is modeled by extracting structural features such as shape and biophysical characteristics of the protein surface from the known protein three-dimensional (3D) structure. The latter extracts sequence information such as amino acid composition (AAC), physicochemical properties, and multiple sequence alignment (MSA) from protein sequences to predict. Due to the structural conserved nature of protein function, the structure-based methods have achieved good results in the early stages. However, since the slow speed of structure determination of wet-lab experiment, it is difficult for these methods to have a great breakthrough in the short term. Despite recent breakthroughs in the field of protein structure prediction by AlphaFold2 [21] and RoseTTAFold [22], the accuracy of predicted 3D structure cannot be fully trusted, especially on orphan proteins that lack high-quality MSA information. In contrast, sequence-based methods only depend on the protein primary sequences, which are abundant and extremely easy to obtain in various protein sequence databases like UniProt [23]. Against this background, it is necessary to develop a reliable and fast sequence-based computational method to accurately predict nucleic acids-binding residue.

A couple dozen of sequence-based methods have been proposed to identify nucleic acids-binding residue. According to prediction targets, these methods can be roughly divided into two categories: methods predicting one of the DNA- or RNA-binding residues only [24] [11] [19], and methods predicting both DNA- and RNA-binding residues [18] [13, 20]. The former focuses on just one of DNA and RNA, and the data sets they used usually contain only DNA-binding protein (DBP) or RNA-binding protein (RBP). For example, PredDBR [24] used the sequence-based cube-format feature and convolutional neural network (CNN) to improve DNA-binding residue prediction performance. Yuan *et al.* first combined the predicted protein structure information by AlphaFold2 and evolution information as feature representation; then, a graph transformer model was employed as DNA-binding residue predictor [11]. In Pprint2 [19], three sequence features, i.e., binary profile, physicochemical properties profile, and evolutionary profile, are utilized as input of CNN model for RNA-binding residue recognition. Agarwal *et al.* [25] implemented a balanced random forest (BRF) classifier with local residue features of RNA-binding residue in protein-RNA complexes as input. In contrast to the above methods, many researchers focus on both DNA- and RNA-binding residues, and they use both DBPs and RBPs to train machine/deep learning models. For example, Wang *et al.* proposed a novel method named iDRNA-ITF which incorporate the functional properties of residues by utilizing an induction and transfer framework for nucleic acid-binding residues identification [18]. In SVMnuc [26], two sequence-based evolution information features, i.e., position-specific scoring matrix (PSSM) [27] and hidden Markov model (HMM) profile [28], and one predicted local structure feature, i.e., protein second structure (SS) [29], are employed and fed to a support vector machine (SVM) [30] model for prediction. Yan *et al.* designed a fast method called DRNApred composed of two logistic regression layers using a comprehensive set of properties of the protein sequence as input for discriminating DNA- and RNA-binding residues [8]. NCBRPred [20] also employed multi-view sequence-based features, i.e., PSSM, HMM, predicted SS and predicted solvent accessibility (SA), as input feature of bidirectional Gated Recurrent Units (BiGRU) [31] for nucleic acid-binding residue identification. These methods have better prediction performance due to the use of both DBP and RBP feature information.

Despite all the above-mentioned methods have made considerable contributions to the development of nucleic acid-binding residue prediction, the challenges remain. First, the prediction results of these methods are highly dependent on the evolutionary information of protein sequences. Since searching for MSAs in a huge protein database is an extremely time-consuming task, they are difficult to generalize to all protein sequences in all organisms. In addition, note that not all proteins can generate high-quality MSAs, and poor MSAs information tends to degrade prediction performance. Second, the number of proteins whose nucleic acid-binding residues have been labelled is very sparse compared to the huge number of protein sequences. As of 13 July 2023, the number of DNA- and RNA-binding protein chains recorded in the BioLip database [32] is only 40,371 and 138,272, respectively, however, the number of protein sequences in UniProtKB/TrEMBL [33] reaches 248,272,897. Most of previous studies utilized only the limited protein information in the training dataset (about 1000 proteins) to learn the identification patterns of nucleic acid-binding residue, which undoubtedly loses the correlation between huge amounts of protein sequences and residues. The nucleic acid-binding residue paradigm is hidden in this vast amount of sequence information.

In this study, we try to mine useful knowledge from as many protein sequences as possible to help in nucleic acid-binding residue recognition, even though some of they are not actually bind to DNA or RNA. Due to the extreme imbalance of positive and negative samples, it is inappropriate to directly construct large imbalanced datasets to train predictive models. Recent major breakthroughs in large-scale protein language models (PLM) such as ESM2 [34] enlightened us. ESM2 was constructed using UniRef50 as a training set (~43 million proteins) by randomly masking and then predicting 15% residues out of protein sequences. This training strategy allows the model to learn the dependencies between residues without labeling. Due to the large number of protein sequences fitted by the trained model and the nature of protein function and structure determined by the sequences, it can be said that the many functional and structural properties of proteins are hidden in the feature representation of the ESM2 model. Based on this idea, here, we propose a fast and accurate method named ESM-NBR, which extract protein feature representations from the ESM2 model and feed them into the multi-task BiLSTM-based neural network model for discriminating DNA- and RNA-binding residues. Experimental results show that the ESM2 feature representation has better performance than the evolutionary information feature of HMM in both of DNA- and RNA-binding residue prediction. Meanwhile, the proposed method demonstrates the MCC values for DNA-binding residues prediction of 0.427 and 0.391 on two independent test sets, i.e., DRNATst-246 and YK17-Tst, which are 18.61 and 10.45% higher than those of the second-best methods, respectively. Moreover, by relying on the graphics processing unit (GPU) for both model inference and feature generation without time-consuming sequence search, ESM-NBR is much faster than previous methods. (~16 times than the second-fastest method on a protein sequence of length 500), which means that ESM-NBR is easily applied to massive amounts of protein sequence data, thus accelerating protein-nucleic acid interaction studies. A user-friendly standalone program and the data of ESM-NBR are freely available for academic use at: https://github.com/pengsl-lab/ESM-NBR.

## II. MATERIALS AND METHODS

### A. Benchmark Datasets

Two pairs of widely used mixed datasets of DBPs and RBPs, i.e., YK17 and DRNA-1314, are employed to evaluate the proposed methods fairly and comprehensively. The detailed composition and description of these two datasets can be found in Supplementary Text S1 at https://github.com/pengsl-lab/ESM-NBR.

### B. Feature representation extraction from ESM2 model

The current mainstream view is generally that protein sequences can determine their 3D structure and thus their function. That is say, at least in theory, the function of a protein (including nucleic acid-binding residue) can be inferred from the primary sequence. However, it has long been a very challenging problem to mine effective discriminatory information from a plethora of protein sequences to accurately predict protein functions. In recent years, the development of transformer-based large-scale language models have led to significant breakthroughs in several field like natural language processing (NLP) [35, 36]. The field of bioinformatics is no exception. The large number of protein sequences provided by high-throughput sequencing technology creates a sufficient knowledge base for protein language model (PLM) [37-39]. ESM2, a large-scale PLM with 15 billion parameters, starts from about 43 million protein primary sequences and learns the mapping relationship among sequences, structures, and functions by masked pre-training, achieving protein tertiary structure prediction with atomic-level accuracy. The prior work of ESM2, i.e., ESM-1b [37], also showed that the feature representation of pre-trained PLM has a significant improvement for protein secondary structure, residue-residue contact, and mutational effects prediction after fine-tuning by supervised learning. Due to the evolutionary conservation and sequence specificity of nucleic acid-binding residues, there is no doubt that the ESM2 feature representation also contains valuable hidden patterns of nucleic acid-protein interactions.

In this study, we used the ESM2 model to generate protein feature representations designed to help improve nucleic acid-binding residue prediction performance. To be specific, we first downloaded ESM2 models with different parameter levels, i.e., 8 million, 35 million, 150 million, 650 million, and 3 billion, according to the tutorial at https://github.com/facebookresearch/esm; then, for each protein sequence $S$ of length $L$ in the datasets, a feature matrix of size $L \times M$ can obtained by feeding $S$ to the ESM2 models, where $M$ means the feature dimension in (320, 480, 640, 1280, 2560) corresponds to models of different sizes. Note that, since the largest ESM2 model has a parameter of 15 billion, which is far beyond the burden of our GPU (Tesla V100 with 16G memory), we did not generate its features. Subsequent experiments demonstrate that the performance of larger model-based feature representations is not necessarily better than that of smaller model. In addition, even on relatively small models, the GPU is still short of memory when dealing with very long sequences such as lengths >1,000. Our solution is to slice the long sequence into multiple relatively short sequences which also have more than 500 amino acids and still retain sufficient sequence contextual information. The well-generated ESM2 feature representations contain important biochemical properties hidden in the protein sequence space rather than just a small amount of knowledge in the training dataset, thus helping to improve nucleic acid-binding residue prediction performance.

### C. Architecture of ESM-NBR

In this study, based on the feature extracted from large protein language model ESM2 and the multi-task BiLSTM-based network, a novel nucleic acid-binding residues prediction method, named ESM-NBR, is proposed and implemented. The overall architecture of ESM-NBR is shown in Figure 1. It is easy to see that the workflow of ESM-NBR can be roughly divided into three steps:

*Step 1*: For each protein primary sequence in the dataset, features containing knowledge of important biochemical attributes are generated by feeding it into the ESM2 model;

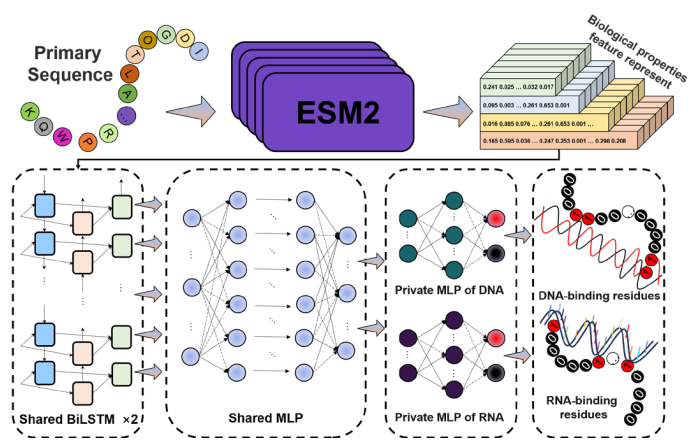

Fig. 1. Architecture of ESM-NBR.

*Step 2*: The well-generated ESM2 feature representation is first inputted into the network shared by DNA- and RNA-binding residues composed of stacked BiLSTM and MLP to learn common knowledge;

*Step 3*: Two private MLP blocks are employed to learn essential authentication information for DNA- and RNA-binding residues identification, respectively. The outputs of the final linear layers are used as the predictive probabilities to determine whether a residue is binding to DNA or RNA.

The multi-task network architecture and training details are described in Supplementary Text S2 at https://github.com/pengsl-lab/ESM-NBR.

### D. Evaluation Indexes

Four evaluation indexes, i.e., Matthew's correlation coefficient (MCC), the area under Receiver Operating Characteristic curve (AUC), the area under Precision-Recall curve (AP), and the area under Cross Prediction Rate-True Prediction Rate (CPR-TPR) curve (AURC) [40], are utilized to assess the proposed ESM-NBR. The detailed description of these four indexes can be found in Supplementary Text S3.

### III. EXPERIMENTAL RESULTS AND ANALYSIS

#### A. Performance comparison of ESM2 feature and evolution information feature

To demonstrate the efficacy of the ESM2 feature representation on nucleic acid-binding residue, the widely used evolutionary information feature of HMM is utilized as a control. Specifically, the HMM feature, ESM2 feature generated by model contained 650 million parameters (Abbreviated as ESM2_650M), and their combination are severally as the input feature of ESM-NBR network for training. The prediction performance on DRNATr-1068 and YK17-Tr over a 10-fold cross validation test are shown in Figure 2 and Table 1. In Table 1, it is obvious that ESM2_650M comprehensively outperforms HMM on both DNA and RNA. Concretely, take results on DRNATr-1068 as an example, the values of MCC, AUC, and AP of ESM2_650M are 0.534, 0.874, and 0.523 (or 0.501, 0.861, and 0.421) considering DNA-binding residue (or RNA-binding residue), which are 90.71, 12.48, and 144.39% (or 187.93, 16.03, and 158.28%) higher than those of HMM respectively. Considering cross-prediction performance, the AURC values of DNA and RNA of ESM2_650M on YK17-Tr are 0.143 and

0.216, which are 148.95 and 115.27% lower than those of HMM separately, indicating ESM2_650 can better distinguish between DNA-binding residue and RNA-binding residue compared to HMM. It is worth that the *p*-values between these two features are so small that the computer cannot calculate it, which means there is a big difference between them. In addition, we can find that the degree of linear correlation between prediction results of these two features is low by visiting PCC values, which further proves the difference between them. By looking at the results of combination feature of ESM2_650M and HMM, the prediction performance does not necessarily get better by simply splicing these two features. For example, the MCC and AP values of DNA-binding residue of ESM2_650M on DRNATr-1068 are 0.534 and 0.523, which are 4.70 and 12.23% higher than those of combination feature respectively. The PCC and *p*-value between ESM2_650M and combination feature show much smaller differences compared to the single HMM. There is no doubt that ESM2_650M dominates in combination feature. Figure 2 shows the PR, CPR-TPR, and ROC curves with corresponding AP, AURC, and AUC values separately. The solid and dashed lines indicate the predicted results for DNA and RNA, respectively. It is intuitive that the PR and ROC curves of HMM are much lower compared to ESM2_650M, and the CPR-TPR curve is much higher especially on RNA-binding residue of YK17-Tr whose AURC is 0.659. Overall, since combination feature consists mainly of ESM2_650M, their curves do not differ much. To sum up, ESM2_650M feature far exceeds the HMM for nucleic acid-binding residue prediction and shows significant difference with it in this section. Since most of the previous methods rely on evolutionary information features heavily, this result provides a new thinking of studying protein-nucleic acid interactions.

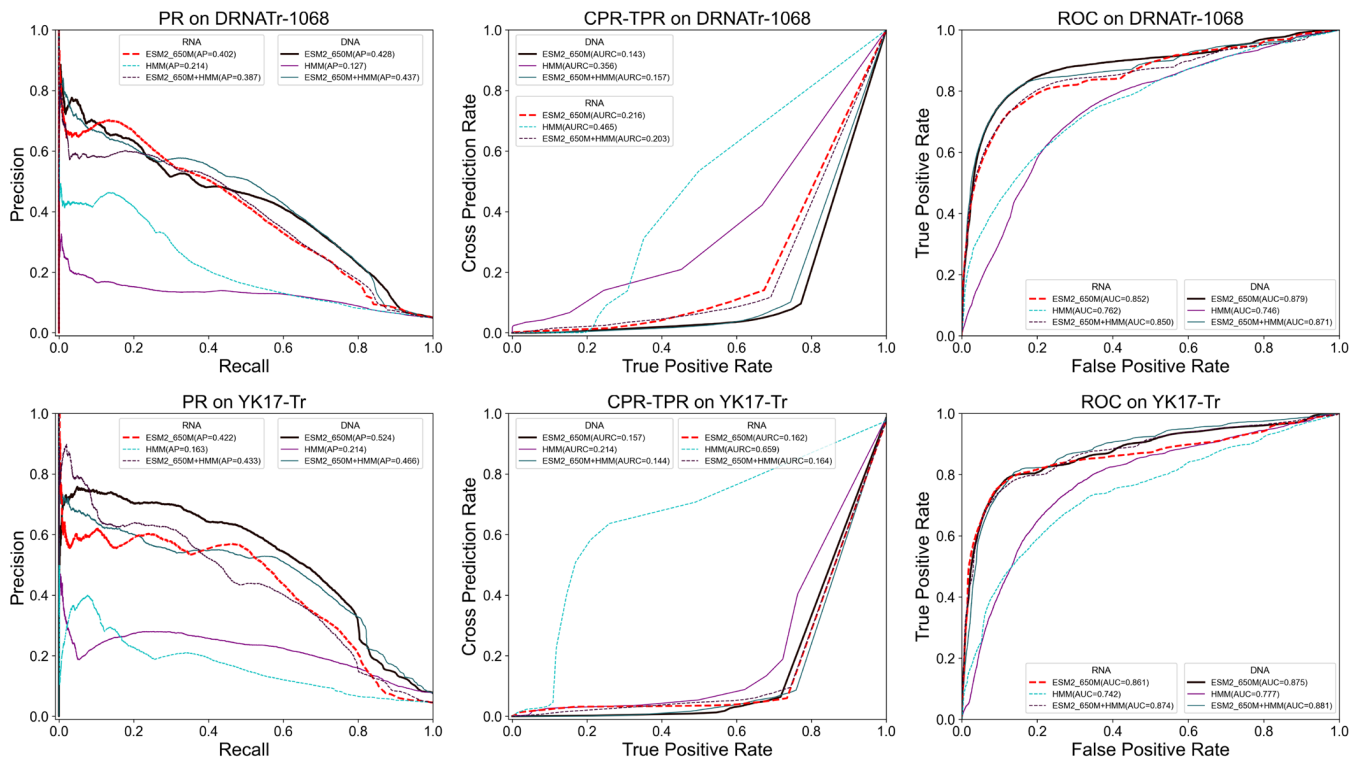

Fig. 2. RP, CPR-TPR, and ROC curves of ESM2_650M, HMM, and their combination on DRNATr-1068 and YK17-Tr over a 10-fold cross validation test. The solid and dashed lines indicate the predicted results for DNA- and RNA-binding residues, respectively. The higher the AUC and AP the better the prediction performance. The lower the AURC the better the prediction performance.

TABLE I PERFORMANCE COMPARISONS OF ESM2_650M AND HMM ON DRNATR-1068 AND YK17-TR OVER A 10-FOLD CROSS VALIDATION TEST

| Dataset | Feature | DNA-binding residue | | | | | | RNA-binding residue | | | | | |
|---|---|---|---|---|---|---|---|---|---|---|---|---|---|
| | | MCC | AUC | AP | AURC | *p*-value [a] | PCC [b] | MCC | AUC | AP | AURC | *p*-value | PCC |
| DRNATr-1068 | ESM2_650M | **0.534** | 0.874 | **0.523** | 0.157 | - | - | **0.501** | 0.861 | 0.421 | **0.162** | - | - |
| | HMM | 0.280 | 0.777 | 0.214 | 0.214 | 4.81e-04 | 3.56e-01 | 0.174 | 0.742 | 0.163 | 0.659 | N/A [c] | 2.94e-01 |
| | Combination | 0.510 | **0.881** | 0.466 | **0.143** | 9.04e-61 | 7.80e-01 | 0.460 | **0.874** | **0.432** | 0.164 | 7.82e-42 | 7.30e-01 |
| YK17-Tr | ESM2_650M | 0.464 | **0.879** | 0.427 | **0.143** [d] | - | - | 0.428 | **0.852** | **0.402** | 0.216 | - | - |
| | HMM | 0.180 | 0.745 | 0.127 | 0.356 | N/A | 3.04e-01 | 0.265 | 0.761 | 0.214 | 0.465 | N/A | 4.92e-01 |
| | Combination | **0.475** | 0.871 | **0.437** | 0.157 | 2.37e-35 | 5.49e-02 | **0.438** | 0.849 | 0.387 | **0.203** | 5.49e-02 | 8.27e-01 |

a. The *p*-values in Student's *t*-test are calculated for the differences between ESM2_650M and other features using the probabilities that each residue is predicted to be a positive sample.
b. The PCC values are calculated for the linear correlation coefficient between ESM2_650M and other features using the probabilities that each residue is predicted to be a positive sample.
c. "N/A" means the value is so small that our computer can't figure it out.
d. Bolded font indicates the best result.

## B. The impact of features generated by ESM2 models with different parameter levels on performance

A number of models with different parameter levels were constructed and compared for extracting important biochemical property knowledge hidden in protein sequence as fully as possible in the study of ESM2 [34]. Since to the extremely complex mapping of protein sequence and structure, the proteins 3D structure at the atomic-level can only be predicted with high accuracy when the number of parameters of the ESM2 model reaches 15 billion. On the nucleic acid-binding residue prediction problem, we do not necessarily use such a large model-generated feature representation as input taking into account possible redundant information. Here, to select features at the appropriate scale, we perform the 10-fold cross validation test on the DRNATr-1068 and YK17-Tr using feature representations generated by ESM2 models with 8 million, 35 million, 150 million, 650 million, and 3 billion parameters on the model in Figure 1, respectively. By looking at Figure 3, it is easy to see that prediction performance is generally poor at lower parameter levels, such as 8 million and 35 million. Obviously small models do not contain enough capacity to learn the vast knowledge of protein sequence space. The features generated by the small model do not contain enough biochemical attributes to accurately predict nucleic acid-binding residues. The overall performance gets progressively better as the number of parameters increases, and the best results are achieved in several indexes when the number of parameters reaches 650 million. For example, in the term of RNA-binding residue prediction, both the MCC and AP values of the feature of model at 650 million level are greater than those of other features, despite the AUC value of the feature of model at 3 billion level is slightly higher than it. Such result suggests that when the number of model parameters is too large, the redundant information in them may lead to deterioration in the performance of the downstream prediction task. By looking at the AURC index marked in purple star, similarly, it reaches lowest on the RNA-binding residue prediction (or DNA-binding prediction) of YK17-Tr (DRNATr-1068) when parameters of ESM2 model at 650 million level. This suggests that DNA-binding residue and RNA-binding residue are well differentiated and that the network learns knowledge about RNA-specificity and DNA-specificity. In conclusion, the feature of ESM2 model at 650 million parameter level is able to predict NBR well in this study, and considering the computational power requirements of larger models, it is appropriate to use it as the input feature of ESM-NBR network.

## C. Fast: highly efficient prediction speed of ESM-NBR

Most existing sequence-based nucleic acid-binding residue prediction methods are relatively inefficient limited by the time-consuming multiple sequence alignment process. ESM-NBR eschews this process altogether, relying solely on large protein language models that have already been well-trained. Here, to clearly demonstrate the advantages of ESM-NBR in the term of prediction efficiency, one protein sequence of length 500 (UniProtKB ID: P17867) is employed to assess the prediction speed of nine existing methods, i.e., DRNAPred, iDRNA-ITF, Pprint2, PST-PRNA, NCBRPred, DNAPred, GraphBind, GraphSite, and PredDBR, and then compared with ESM-NBR. With the exception of prediction of GraphBind, which come from a standalone running package program, the results of other eight predictors are obtained from their web services. Taking into account network fluctuations and server busyness, the prediction times of these methods are the lowest values selected from the results of multiple tests. The prediction time of ESM-NBR includes the whole process of model loading to ESM2 feature generation to model inference. A detailed comparison is shown in Figure 4A. It is obvious that the ESM-NBR has the shortest prediction time of 5.52 seconds, which is about 16 times faster than the second fastest method, i.e., DRNAPred. The prediction times of the other methods ranged from 130 seconds to 8,220 seconds, which is hardly in the same order of magnitude as ESM-NBR. Most of these methods require multiple sequence alignment to generate evolutionary information-based files first like PSSM and HMM, or some require local or global protein structures to be predicted first such as GraphSite, which is main reason that limits their predictive efficiency. Figure 4B shows the prediction time of ESM-NBR on protein sequences of different lengths ranged from 100 and 5,000 using CPU and GPU, respectively. All results were obtained by independently repeating the experiment 10 times and then averaging the times. By observing Figure 4B, on both two devices, the prediction time increases as the length of the sequence increases. However, even for sequences of length 5,000, the maximum prediction time is less than 2 minutes. It is worth noting that the time tested with the GPU is higher than that of with the CPU. Limited by the GPU memory size, only one sequence is tested serially at a time, thus not taking advantage of GPU parallel computing. The large-scale prediction time can be further optimized if the GPU has enough memory. Although GPU with large amounts of memory is not always available for most researchers, the computational speed of ESM-NBR on CPU is also much better than existing methods.

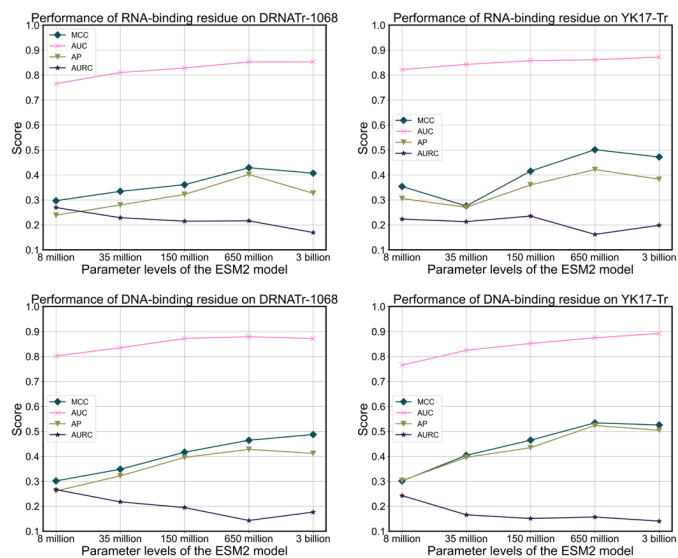

Fig. 3. Prediction performance changes of features generated by ESM2 models with different parameter levels on DRNATr-1068 and YK17-Tr over a 10-fold cross validation test.

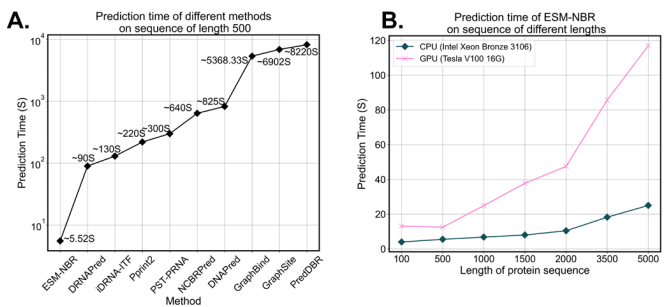

Fig. 4. Comparison of prediction times. (A): prediction times of ESM-NBR and existing methods on a sequence of length 500; (B): prediction times of ESM-NBR on protein sequences of different lengths using GPU and CPU respectively.

As a conclusion, the efficient and rapid prediction of ESM-NBR can be easily applied to massive protein sequences, thus accelerating nucleic acid-protein interaction studies.

### D. Accurate: prediction performance comparison with the state-of-the-art prediction methods

To evaluate the prediction performance of ESM-NBR, four state-of-the-art nucleic acid-binding residues predictors, i.e., iDRNA-ITF, DRNAPred, GraphBind, and NCBRPred, are employed as control. To ensure the fairness, the prediction results of ESM-NBR on these two test sets are obtained from the model trained by YK17-Tr and DRNATr-1068, respectively. The prediction results of iDRNA-ITF, DRNAPred, and NCBRPred are generated by feeding the protein sequences to their web servers. Since the web server of GraphBind cannot accept multiple sequences at once, we downloaded the source code provided in http://www.csbio.sjtu.edu.cn/bioinf/GraphBind/sourcecode.html and run it. It is noted that GraphBind is a structure-based method. There is no way for us to directly use the structure of YK17-Tst to make predictions because the proteins in the YK17-Tst contain disordered regions of unknown structure. Here, the predicted 3D structures of AlphaFold2 (ColabFold [41] version with default parameter) are employed as the alternative to the real 3D structure as the input to the GraphBind.

The detailed comparisons on DRNATst-246 and YK17-Tst are listed in Table 2. By looking at MCC values of DNA-binding residue prediction, ESM-NBR achieves the best results of 0.427 and 0.391 on both datasets, which are 32.60, 284.68, 18.61, and 85.65%, and 10.45, 90.73, 26.12, and 26.96% higher than iDRNA-ITF, DRNAPred, GraphBind, and NCBRPred on DRNATst-246 and YK17-Tst separately. In the term of cross-prediction, the AURC values of ESM-NBR on two datasets for DNA-binding residue prediction are 0.121 and 0.195, which is second only to DRNAPred and better than the other three methods. Not lost on us is the fact that AURC values of DRNAPred for predicting RNA-binding residues are 0.937 and 0.696, indicating that DRNAPred predicates most of the native RNA-binding residues as DNA-binding residues. Potentially the biggest reason should be that DRNAPred model overfitted the DNA-binding residue data during the training stage. Turning the attention to the RNA-binding residue prediction, the AUC values of ESM-NBR on DRNATst-246 is 0.838, which is 17.36, 71.37, 0.47, and 20.57% higher than other methods respectively. On YK17-Tst, ESM-NBR also gets the second-best prediction behind iDRNA-ITF. Due to the sheer number of all the residues in test sets, we use the probability that the native DNA-binding (or RNA-binding) residues are predicted to be positive samples to calculate the $p$-values. It is clear that the $p$-values between ESM-NBR and most of the other methods is very small, demonstrating a statistically significant difference. Also, the PCC values mean a lower degree of linear correlation.

For a more comprehensive comparison, we calculate MCC values for DNA-binding (or RNA-binding) residues prediction individually for each protein on both test sets and then plot scatter plots shown in Figure 5. Figures 5A~5D are the comparison results on DRNATst-246 by separating DBP and RBP. Figures 5E~5H are the comparison results on the YK17-Tst by calculating MCC values of DNA/RNA-binding

TABLE II  PERFORMANCE COMPARISONS OF THE STATE-OF-THE-ART METHODS AND ESM-NBR ON DRNATST-246 AND YK17-TST OVER INDEPENDENT VALIDATION

| Dataset | Model | DNA-binding residue | | | | | | RNA-binding residue | | | | | |
|---|---|---|---|---|---|---|---|---|---|---|---|---|---|
| | | MCC | AUC | AP | AURC | $p$-value [a] | PCC [b] | MCC | AUC | AP | AURC | $p$-value | PCC |
| DRNATst-246 | iDRNA-ITF | 0.322 | 0.839 | 0.303 | 0.318 | 3.65e-29 | 5.37e-01 | 0.207 | 0.714 | **0.247** [c] | 0.335 | 7.18e-68 | 3.87e-01 |
| | DRNAPred | 0.111 | 0.686 | 0.073 | **0.087** | 3.39e-146 | 2.37e-01 | 0.019 | 0.489 | 0.029 | 0.937 | 4.01e-93 | 5.88e-02 |
| | GraphBind | 0.360 | **0.922** | 0.322 | 0.454 | 1.66e-25 | 4.71e-01 | **0.221** | 0.834 | 0.128 | 0.471 | 4.66e-43 | 2.98e-01 |
| | NCBRPred | 0.230 | 0.819 | 0.178 | 0.229 | 8.15e-220 | 3.91e-01 | 0.155 | 0.695 | 0.119 | 0.594 | 7.42e-62 | 2.12e-01 |
| | ESM-NBR | **0.427** | 0.901 | **0.405** | 0.121 | - | - | 0.218 | **0.838** | 0.148 | 0.462 | - | - |
| YK17-Tst | iDRNA-ITF | 0.354 | **0.881** | **0.381** | 0.325 | 1.43e-05 | 5.54e-01 | **0.339** | **0.870** | **0.285** | **0.162** | 4.24e-100 | 2.87e-01 |
| | DRNAPred | 0.205 | 0.767 | 0.194 | **0.038** | 2.65e-05 | 3.80e-01 | 0.122 | 0.670 | 0.096 | 0.696 | 2.08e-13 | 1.63e-01 |
| | GraphBind [d] | 0.310 | 0.866 | 0.291 | 0.509 | 9.162e-36 | 4.70e-01 | 0.185 | 0.757 | 0.134 | 0.481 | 9.80e-01 | 1.74e-01 |
| | NCBRPred | 0.308 | 0.840 | 0.304 | 0.242 | 5.22e-15 | 4.88e-01 | 0.217 | 0.767 | 0.192 | 0.474 | 3.07e-09 | 2.82e-01 |
| | ESM-NBR | **0.391** | **0.881** | 0.350 | 0.195 | - | - | 0.276 | 0.785 | 0.232 | 0.462 | - | - |

a. The $p$-values in Student's $t$-test are calculated for the differences between other methods and ESM-NBR using the probabilities that the native nucleic acid-binding residue is predicted to be a positive sample.
b. The PCC are calculated for the linear correlation coefficient between other methods and ESM-NBR model using the probabilities that each residue is predicted to be a positive sample.
c. Bolded font indicates the best result.
d. Since the proteins in the YK17-Tst contain disordered regions of unknown structure, the protein structures used for prediction results of GraphBind are from AlphaFold2.

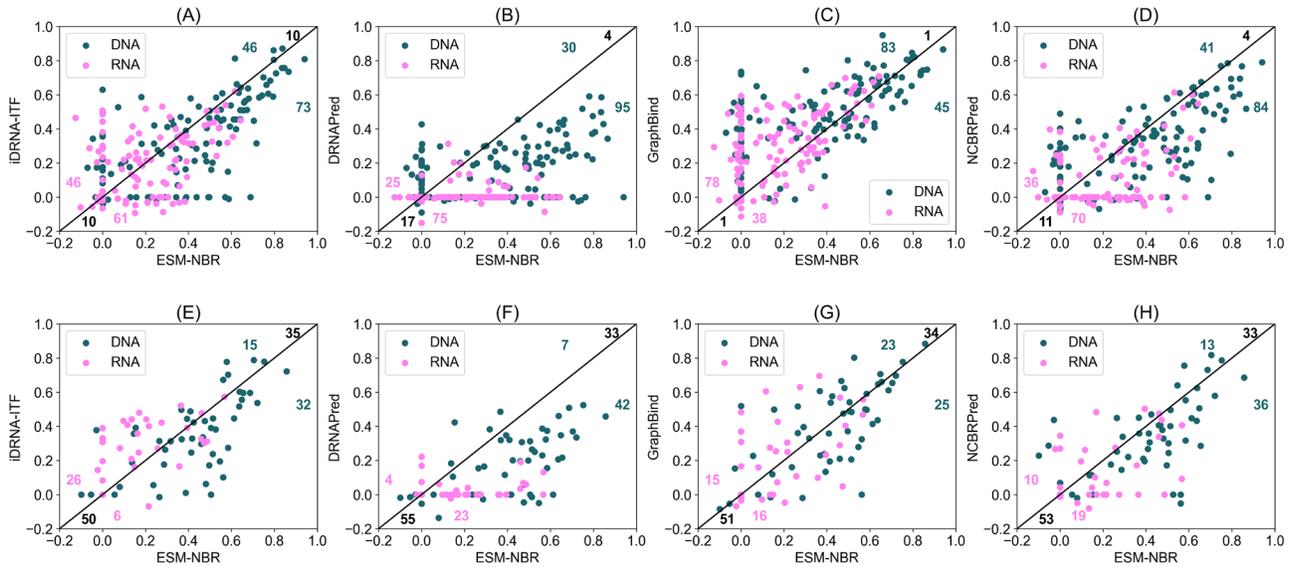

Fig. 5. Head-to-head comparisons of MCC values of ESM-NBR and four state-of-the-art nucleic acid-binding residue prediction methods on DRNATst-246 and YK17-Tst. The green (or pink) numbers on the diagram indicate the number of DBPs (or RBPs) which locate in the upper or lower triangle. The black numbers located next to the black slashes indicate the number of DBPs (or RBPs) with equal MCC values for both two methods. (A~D): comparison results on DRNATst-246; (E~H): comparison results on YK17-Tst. Since YK17 contains proteins that bind both DNA and RNA simultaneously, instead of splitting the dataset into DNA and RNA subsets, we calculate MCC values for DNA- and RNA-binding residues separately for each protein in YK17. In contrast, DRNATst-246 is segmented into DNA and RNA subsets to calculate MCC values for DNA- and RNA-binding residues, respectively.

residues for each protein considering that some of the proteins in YK17-Tst bind both DNA and RNA. It is clear that ESM-NBR outperforms most other methods on both DNA- and RNA-binding residues prediction. Specifically, take result on DRNATst-246 as an example, out of 129 DNA-binding proteins (or 117 RNA-binding proteins), there are 73, 95, 45, and 84 (or 61, 75, 38, and 70) proteins which ESM-NBR has higher MCC values than other methods, respectively. In addition, we find that the pink dots are more towards the bottom left compared to the green dots, meaning all methods predicted poorly on RNA-binding residue than on DNA-binding residue even on DRNATst-246 whose training data set has more RNA-binding residues than DNA-binding proteins. The potential reason may be the simplicity of the pattern of DNA-binding residue relative to that of RNA-binding residues. As a conclusion, the outstanding performance of ESM-NBR surpasses most of the existing methods and shows a clear differentiation from them.

## IV. CONCLUSIONS

Accurate identification of nucleic acid-binding residue is a key step in understanding nucleic acid-protein interactions. In this study, to enhance the performance and speed of nucleic acid-binding residues identification, a new sequence-based method called ESM-NBR is designed and implemented. In ESM-NBR, the protein primary sequence is first fed to the large protein language model, i.e., ESM2, to generate important biological properties feature representation; then, the generated feature is inputted to the stacked BiLSTM and MLP layers shared by DNA- and RNA-binding residues; finally, two private MLP blocks are used to discriminate DNA- and RNA-binding residues, respectively. Experimental results on benchmark test data sets demonstrate that both the prediction speed and accuracy of ESM-NBR exceeds most of the state-of-the-art nucleic acid-binding residue prediction methods. The standalone program, supplementary information and data of ESM-NBR are available for free at https://github.com/pengsl-lab/ESM-NBR.

In the future study, to further improve the performance of nucleic acid-binding residues prediction, the following points will be the focus of our research: (1) collecting large numbers of nucleic acid-bind protein sequences to pre-train large-scale protein language models; (2) applying large-scale protein language models to a variety of downstream tasks such nucleic acid-binding protein prediction, nucleic acid-binding residue prediction, and protein-nucleic acid complex structure prediction; (3) using transfer learning algorithm to import the helpful knowledge from proteins binding to other ligands such as ATP. In spite of ESM-NBR still has room for improvement, it should be a powerful tool in the field of the nucleic acid-bind residue prediction.

## DATA AVAILABILITY

A user-friendly standalone program, supplementary information and the data of ESM-NBR are freely accessible at https://github.com/pengsl-lab/ESM-NBR. The supplementary information includes a detailed description of the data set, neural network architecture, and evaluation indexes used in this study. In addition, comparative experiments with single-task and multi-task models, as well as case studies, are also performed.

## AKNOWLEDGMENT

This work was supported by National Key R&D Program of China 2022YFC3400400; NSFC Grants U19A2067; Key Technologies R&D Program of Guangdong Province